\documentclass[]{emulateapj}
\usepackage{epsfig}

\usepackage{multirow}

\usepackage{rotating}
\usepackage{natbib}
\usepackage{latexsym}
\bibpunct{(}{)}{;}{a}{}{,}
\usepackage{url}

\begin{document}

\newcommand{\ledd}{%
$L_{Edd}$}

\def\rem#1{{\bf (#1)}}
\def\hide#1{}

%\title{Thermonuclear flashes from the neutron star X-ray binary Circinus X-1:\\ The return of the bursts}
\title{The return of the bursts:\\ Thermonuclear flashes from Circinus X-1}

\author{M. Linares\altaffilmark{1,2}, A. Watts\altaffilmark{3}, D. Altamirano\altaffilmark{3}, P. Soleri\altaffilmark{4}, N. Degenaar\altaffilmark{3}, Y. Yang\altaffilmark{3}, R. Wijnands\altaffilmark{3}, P. Casella\altaffilmark{5}, J. Homan\altaffilmark{1}, D. Chakrabarty\altaffilmark{1}, N. Rea\altaffilmark{6}, M. Armas-Padilla\altaffilmark{3}, Y. Cavecchi\altaffilmark{3,7}, M. Kalamkar\altaffilmark{3}, R. Kaur\altaffilmark{3}, A. Patruno\altaffilmark{3}, M. van der Klis\altaffilmark{3}}

\date{}

\altaffiltext{1}{Massachusetts Institute of Technology - Kavli
Institute for Astrophysics and Space Research, Cambridge, MA 02139,
USA}

\altaffiltext{2}{Rubicon Fellow}

\altaffiltext{3}{Astronomical Institute ``Anton Pannekoek'', University of Amsterdam and Center for High-Energy Astrophysics, PO BOX 94249, 1090 GE, Amsterdam, Netherlands}

\altaffiltext{4}{Kapteyn Astronomical Institute, University of Groningen, PO BOX 800, 9700 AV, Groningen, The Netherlands}

\altaffiltext{5}{School of Physics and Astronomy, University of Southampton, Southampton, Hampshire, SO17 1BJ, United Kingdom}

\altaffiltext{6}{Institut de Ciencies de l'Espai (ICE, CSIC--IEEC),
Campus UAB, Facultat de Ciencies, Torre C5-parell, 2a planta, 08193,
Bellaterra (Barcelona), Spain}

\altaffiltext{7}{Sterrewacht Leiden, University of Leiden, Niels Bohrweg 2, 2333 CA, Leiden, The Netherlands}


\begin{abstract}

We report the detection of 15 X-ray bursts with {\it RXTE} and {\it
  Swift} observations of the peculiar X-ray binary Circinus~X-1 during
its May 2010 X-ray re-brightening. These are the first X-ray bursts
observed from the source after the initial discovery by Tennant and
collaborators, twenty-five years ago. By studying their spectral
evolution, we firmly identify nine of the bursts as type I
(thermonuclear) X-ray bursts. We obtain an arcsecond location of the
bursts that confirms once and for all the identification of Cir~X-1 as
a type~I X-ray burst source, and therefore as a low magnetic field
accreting neutron star. The first five bursts observed by {\it RXTE}
are weak and show approximately symmetric light~curves, without
detectable signs of cooling along the burst decay. We discuss their
possible nature. Finally, we explore a scenario to explain why Cir~X-1
shows thermonuclear bursts now but not in the past, when it was
extensively observed and accreting at a similar rate.

\end{abstract}

\maketitle

\section{Introduction}
\label{sec:intro}

Discovered during the early years of X-ray astronomy \citep{Margon71}
and frequently observed ever since, the peculiar X-ray binary
Cir~X-1 was initially classified as a black hole
candidate (BHC), due to spectral and variability similarities to
Cyg~X-1 \citep{Toor77}. Its $\sim$16.6 day period, discovered in the
X-ray band by \citet{Kaluzienski76} and observed at different epochs
and wavelengths, is attributed to enhanced accretion near periastron
passage in a highly eccentric orbit (\citealt{Murdin80}; see also
\citealt{Jonker07}).

%(very soft and very hard states)
%(rapid flickering) 

Thermonuclear explosions on the surface of accreting neutron stars
\citep[type I X-ray bursts; see, e.g.,][for a recent
  review]{Strohmayer06} are one of the few signatures that allow us to
identify unambiguously a compact object as a neutron star (NS). 
%We
%know more than 90 low-mass X-ray binaries (LMXBs) that have shown type
%I X-ray bursts \citep[see, e.g., ][]{Galloway08}.
%
The defining observational property of type I X-ray bursts is a mainly
thermal spectrum with black body temperature that decays along the
burst tail \citep[``cooling tail''; e.g.][]{Lewin93}.
In 1984--1985, 11 X-ray bursts were discovered in {\it EXOSAT}
observations of Cir~X-1 \citep{Tennant86a,Tennant86b}. Three of these
could be identified as type I X-ray bursts based on their cooling
tails \citep{Tennant86b}, while a type II identification \citep[see
  e.g.][]{Lewin93} could not be discarded for the remaining eight
bursts. The discovery of type I X-ray bursts led to the conclusion
that the compact object in Cir~X-1 is a NS \citep{Tennant86b}.
Since then many X-ray missions have observed Cir~X-1 but no X-ray
bursts were detected \citep[see, e.g., ][ for a search of 2.7
  Msec of Cir~X-1 data from {\it RXTE}]{Galloway08}.

Despite the BHC similarities, \citep[][which are not unique to this
  source]{Toor77,Klis94}, the fast X-ray variability of Cir~X-1 has
been described as similar to the most luminous NS-LMXBs (Z-sources)
and the less luminous NS-LMXBs (atoll sources) \citep[see,][and
  references therein]{Oosterbroek95,Shirey98,Soleri09}. The complex
phenomenology observed in this source, together with the lack of
coherent pulsations and the fact that no X-ray bursts have been
detected since 1985, have led to the speculation that the compact
object in Cir~X-1 may be a BH, and that the X-ray bursts discovered by
\citet{Tennant86a,Tennant86b} came from a nearby system in the {\it
  EXOSAT} $\sim$0.75$^\circ$ FoV. The most recent indication that
Cir~X-1 is indeed a NS is the detection of twin kilohertz
quasi-periodic oscillations \citep[kHz QPOs;
][]{Boutloukos06}. Although this phenomenon can be considered a proof
of the NS nature of the accretor \citep{vanderKlis06}, the very
peculiar properties of these kHz QPOs prevented a conclusive consensus
on the identification of the compact object in Cir~X-1.

\begin{figure*}[ht]
\centering
%  \begin{center}
  \resizebox{0.95\columnwidth}{!}{\rotatebox{-90}{\includegraphics[]{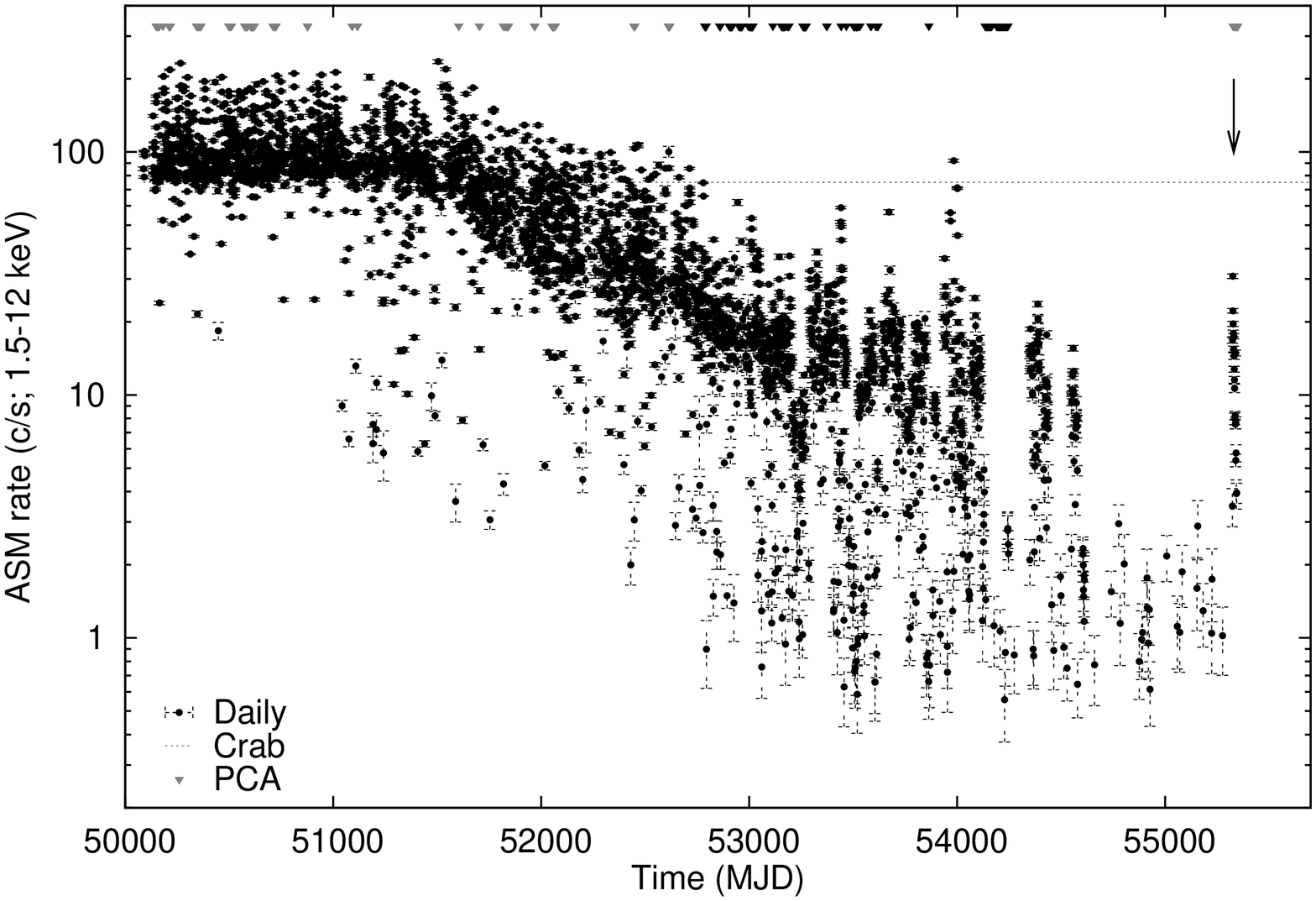}}}
%  \caption{}
  \resizebox{0.9\columnwidth}{!}{\rotatebox{-90}{\includegraphics[]{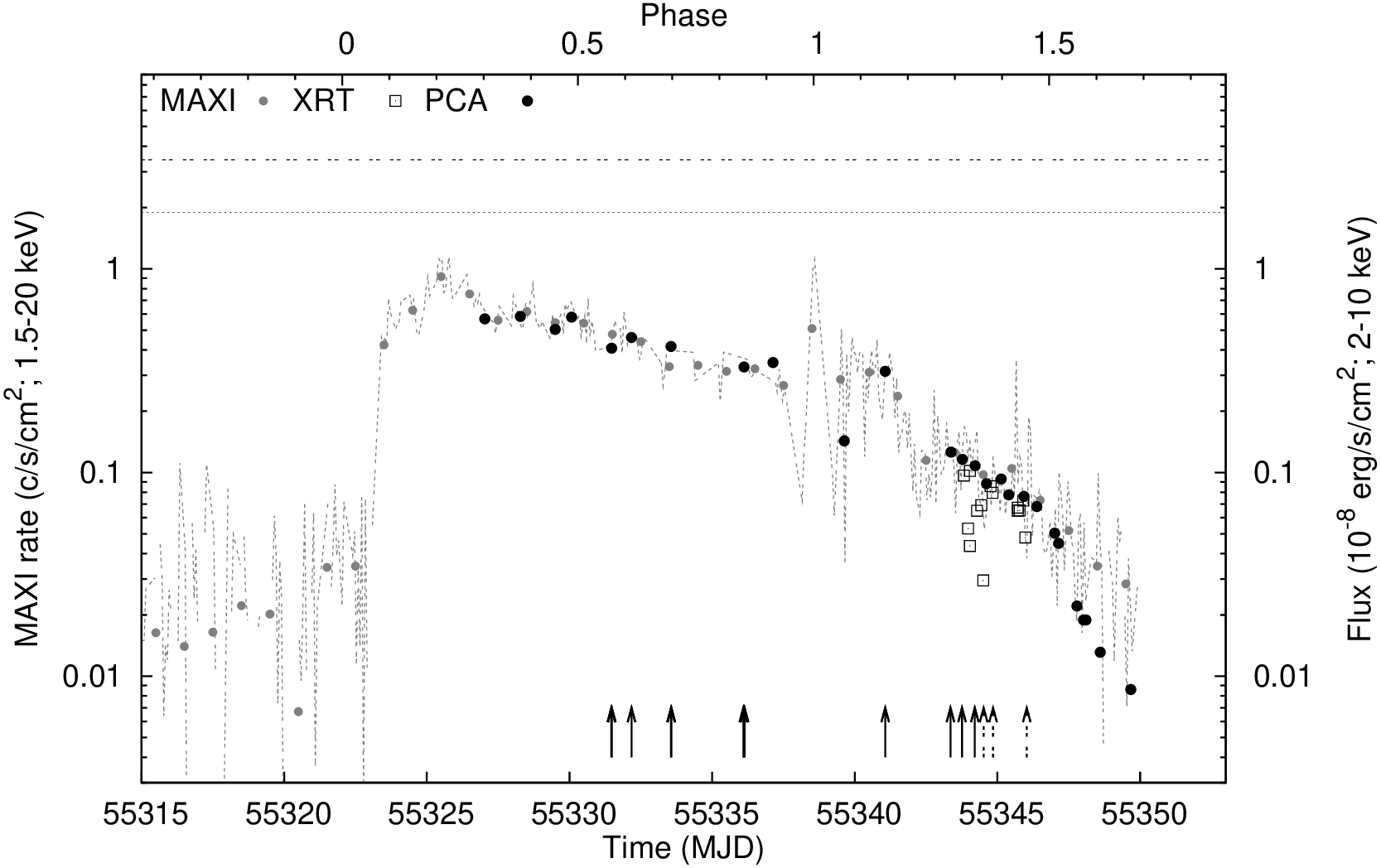}}}
  \caption{
{\it Left:} 1996--2010 {\it RXTE}-ASM \citep{Levine96} light~curve of Cir~X-1, showing daily average count rates (filled black circles) in the 1.5--12~keV band (only detections above a 3~$\sigma$ threshold are displayed). The horizontal dotted line shows the approximate count rate of the Crab, for comparison. Triangles along the top axis mark times of all pointed {\it RXTE}-PCA observations of Cir~X-1; black triangles show observations taken between April 2003 and April 2010, when the ASM rate was similar to that measured during May 2010 but no bursts were detected. The arrow shows the epoch of the X-ray bursts reported in this work.
{\it Right:} Combined {\it RXTE}-PCA \citep{Jahoda96} and {\it MAXI}-GSC \citep{Matsuoka09} light~curve during the May 2010 re-brightening and bursting period. Shown are single orbit and daily average 1.5--20~keV {\it MAXI} count rates (dashed grey line and filled grey circles, respectively), and the 2--10~keV absorbed flux measured by {\it RXTE}--PCA and {\it Swift}-XRT (filled black circles and open squares, respectively). The times of {\it RXTE} and {\it Swift} X-ray bursts are marked with solid and dashed black arrows, respectively. The grey dashed arrow shows the time of the X-ray burst detected by {\it MAXI} (http://maxi.riken.jp/news/en/?p=503). The Eddington flux (for an Eddington luminosity of 2.5$\times$10$^{38}$~erg~s$^{-1}$) at 10.5 and 7.8~kpc \citep{Jonker04} is shown with a horizontal dotted and dashed line, respectively. Orbital phase is shown on the top axis \citep[from radio ephemeris, where phase zero corresponds to onset of radio flares and is associated to periastron passage]{Nicolson07}.}
    \label{fig:lc2}
%\epsscale{1.0}
% \end{center}
\end{figure*}
%path is: /scratch/mlinares/rxte/CirX-1/

In this Letter, we report the detection of 15 X-ray bursts from {\it
  RXTE} and {\it Swift} observations of Cir~X-1. As noted by
\citet{Linares10a}, these are the first X-ray bursts observed from
Cir~X-1 after the initial discovery, 25 years ago
\citep[][]{Tennant86a,Tennant86b}. We identify 9 of them as
thermonuclear -type I- X-ray bursts and obtain an arc-second location
fully consistent with the position of Cir~X-1, unambiguously
identifying Cir~X-1 as a low magnetic field accreting neutron star. We
discuss the nature of the bursts and the burning regime, as well as
different scenarios that may explain the change in their
properties. Finally, we address the question of why Cir~X-1 has not
shown type I X-ray bursts in the last 25 years, and why it shows them
now.

\section{Data Analysis and Results}
\label{sec:dataresults}

On May 7th, 2010, {\it MAXI}-GSC observations detected an X-ray
re-brightening of Cir~X-1 after two years of very low flux
\citep[below $\sim$70~mCrab and with an average flux of
  $\sim$10~mCrab,][see also Fig.~\ref{fig:lc2},
  right]{Nakajima10}. Monitoring observations with {\it RXTE} started
on May 11th, 2010.

\subsection{RXTE}
\label{sec:rxte}

\begin{figure}[h!]
  \begin{center}
  \resizebox{0.8\columnwidth}{!}{\rotatebox{-90}{\includegraphics[]{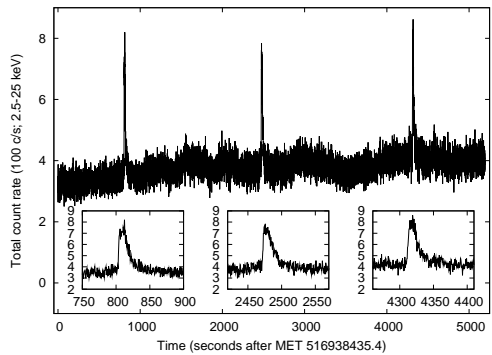}}}
  \caption{PCA 0.5s time resolution light~curve of the three consecutive bursts detected on May 20, 2010 (R6-R8). Insets show zoomed 150~s-long light~curves of each burst.}
    \label{fig:triple}
%\epsscale{1.0}
 \end{center}
\end{figure}

We performed a search for X-ray bursts in all (26) {\it RXTE}-PCA
observations of Cir~X-1 taken between May 11th and June 1st, 2010,
using 1~s time resolution PCA light~curves (Standard~1 mode; full
2--60keV energy band). We find a total of 12 X-ray bursts that we
label in the present work as R1-R12 (the first 9 were reported in
\citealt{Linares10a}; see also \citealt{Papitto10a}). Two pairs of
consecutive X-ray bursts were detected on May 15 and May 17, with wait
times of $\sim$20 minutes. On May 20 {\it RXTE} observed three
consecutive type I X-ray bursts (shown in Figure ~\ref{fig:triple})
with slightly longer wait times ($\sim$28 and $\sim$31 minutes).

In order to study their spectral evolution, we performed time-resolved
spectroscopy of all the X-ray bursts. We extracted PCA spectra in 1~s
steps using event mode data (E\_125us\_64M\_0\_1s) of all layers and
available PCUs, grouped the resulting spectra to have a minimum of 15
counts per bin and added a 1\% systematic error to all channels. For
each 1-s spectrum, we corrected for deadtime following the PCA team
directions\footnote{\url{http://heasarc.gsfc.nasa.gov/docs/xte/recipes/pca\_deadtime.html}},
and created a response matrix using {\it pcarsp} (v. 11.7). Following
standard approach in X-ray burst analysis \citep[e.g.][]{Kuulkers02},
we subtracted a 100-s pre-burst spectrum to account for background,
non-burst emission (which includes instrumental background and
accretion flux). We use the same pre-burst spectrum to measure the
0.5-50~keV persistent flux prior to each burst, as a proxy for the
mass accretion rate at the time of ignition. We fitted each 1-s burst
spectrum in the 2.5--25.0~keV energy range within XSPEC (v. 12.5) with
an absorbed black body model \citep[wabs*bbodyrad, with abundances
  from][]{Anders82}. Given that the PCA bandpass is not well suited
for constraining the absorbing column density, we fixed this parameter
to the value found by \citet{Iaria01}, 1.6$\times$10$^{22}$cm$^{-2}$
\citep[absorption can be substantially higher near periastron passage,
  see e.g.][but none of the bursts analyzed herein occurred around
  phase 1]{Schulz08}. All luminosities, energies and black-body radii
presented in this work use a fiducial distance of 7.8~kpc, yet the
range of distance estimates must be considered
(e.g. \citealt{Jonker04} find a distance to Cir~X-1 between 7.8 and
10.5~kpc based on the type I X-ray bursts from \citealt{Tennant86b}).

%path is: /scratch/mlinares/rxte/CirX-1/RT/BURSTS/

%preprint:
%\begin{table}[b]
%apjlet:
\begin{table*}[ht]
%\rotate
%\scriptsize
\tiny
\caption{Properties of the 15 X-ray bursts from Cir~X-1 analyzed in this work.}
\begin{minipage}{\textwidth}
\centering
\begin{tabular}{ c c c c c c c c c c c}
\hline\hline

ID & OBS.\footnote{Observation ID, preceded by 95422-01- and 00030268- in the {\it RXTE} and {\it Swift} observations, respectively. XRT observing mode between parenthesis.}  & Peak time      & Orbital & Net peak\footnote{Persistent level subtracted; 2.5--25.0~keV and 0.5--10.0~keV energy range for R1-R12 and S1-S3, respectively. Number of active PCUs indicated between brackets in the {\it RXTE} bursts.}  & Rise\footnote{Defined as time from 25\% to 90\% of peak flux \citep[see e.g.][]{Galloway08}.}  & L$_{peak}$\footnote{Bolometric burst peak luminosity and total energy, after subtracting persistent emission. Persistent luminosity in the 0.5--50~keV band. All use a distance of 7.8~kpc \citep[Sec.~\ref{sec:rxte}]{Jonker04}.} & Energy$^d$ & L$_{pers}$$^d$ & Wait\footnote{Only quoted when two or more consecutive bursts are detected. Alpha parameter is defined as L$_{pers}$ times wait time divided by total burst energy.}  & $\alpha$$^e$ \\
   & ID &          (MJD) &   phase   &rate (c/s) & time (s) & (10$^{37}$erg~s$^{-1}$) & (10$^{38}$erg) &  (10$^{37}$erg~s$^{-1}$) & time (min) &  \\
\hline
\multicolumn{11}{c}{{\it RXTE}}\\
\hline
R1 & 02-03 & 55331.46840 & 0.570    & 153.1 [1] & 9.6 $\pm$ 0.5 & 1.3 $\pm$ 0.9 & 1.3 $\pm$ 0.6 & 4.6 $\pm$ 0.3 & - & - \\
R2 & 02-03 & 55331.48227 & 0.572    & 220.2 [1] & 7.3 $\pm$ 0.5 & 2.0 $\pm$ 0.8 & 3.5 $\pm$ 1.7 & 4.4 $\pm$ 0.2 & 20.0 & 129 \\
R3 & 02-02 & 55332.17123 & 0.61     & 153 [1] & 8.8 $\pm$ 0.5 & 1.3 $\pm$ 0.8 & 1.5 $\pm$ 0.7 & 5.6 $\pm$ 0.3 & - & - \\
R4 & 02-00 & 55333.55562 & 0.697    & 279.1 [1] & 7.5 $\pm$ 0.5 & 2.5 $\pm$ 0.8 & 3.7 $\pm$ 1.6 & 4.7 $\pm$ 0.3 & - & - \\
R5 & 02-00 & 55333.56971 & 0.698    & 241.7 [1] & 10.2 $\pm$ 0.5 & 2.2 $\pm$ 0.9 & 2.4 $\pm$ 1.1 & 5.3 $\pm$ 0.3 & 20.2 & 238 \\
R6 & 02-04 & 55336.09307 & 0.851    & 453.1 [1] & 2.0 $\pm$ 0.5 & 4.2 $\pm$ 0.8 & 5.8 $\pm$ 1.9 & 4.2 $\pm$ 0.2 & - & - \\
R7 & 02-04 & 55336.11231 & 0.852  & 393.9 [1] & 2.2 $\pm$ 0.5 & 3.6 $\pm$ 0.9 & 5.3 $\pm$ 1.8 & 4.0 $\pm$ 0.2 & 27.7 & 87 \\
R8 & 02-04 & 55336.13367 & 0.853    & 453.9 [1] & 5.3 $\pm$ 0.5 & 4.1 $\pm$ 0.9 & 5.5 $\pm$ 1.9 & 4.6 $\pm$ 0.2 & 30.8 & 152 \\
R9 & 03-02 & 55341.06464 & 1.15    & 724.5 [2] & 5.1 $\pm$ 0.5 & 3.5 $\pm$ 0.6 & 11 $\pm$ 3.8 & 4.8 $\pm$ 0.2 & - & - \\
R10 & 03-03 & 55343.35345 & 1.29    & 795.7 [2] & 2.5 $\pm$ 0.5 & 3.8 $\pm$ 0.5 & 7.6 $\pm$ 1.8 & 1.7 $\pm$ 0.1 & - & - \\
R11 & 03-04 & 55343.75671 & 1.31    & 937.4 [2] & 1.8 $\pm$ 0.5 & 4.4 $\pm$ 0.5 & 8.6 $\pm$ 2.0 & 1.6 $\pm$ 0.1 & - & - \\
R12 & 04-00 & 55344.20388 & 1.34    & 1057.7 [2] & 1.8 $\pm$ 0.5 & 4.5 $\pm$ 0.5 & 9.7 $\pm$ 2.5 & 1.5 $\pm$ 0.1 & - & - \\
\hline
\multicolumn{11}{c}{{\it SWIFT}}\\
\hline
S1 & 032(WT) & 55344.50640 & 1.36 & $\sim$15 & $\sim$1 & - & - & - & - & - \\
S2 & 033(WT) & 55344.84431 & 1.38 & $\sim$35 & $\sim$1 & - & - & - & - & -  \\
S3 & 034(PC) & 55346.02417 & 1.45 & $\gtrsim$15 & $\lesssim$2.5 & - & - & - & - & -  \\
\hline\hline
\end{tabular}
\end{minipage}
\label{table:rbursts}
%preprint:
%\end{table}
%apjlet:
\end{table*}

\begin{figure}[ht]
%  \begin{center}
\centering
  \resizebox{1.06\columnwidth}{!}{\rotatebox{0}{\includegraphics[]{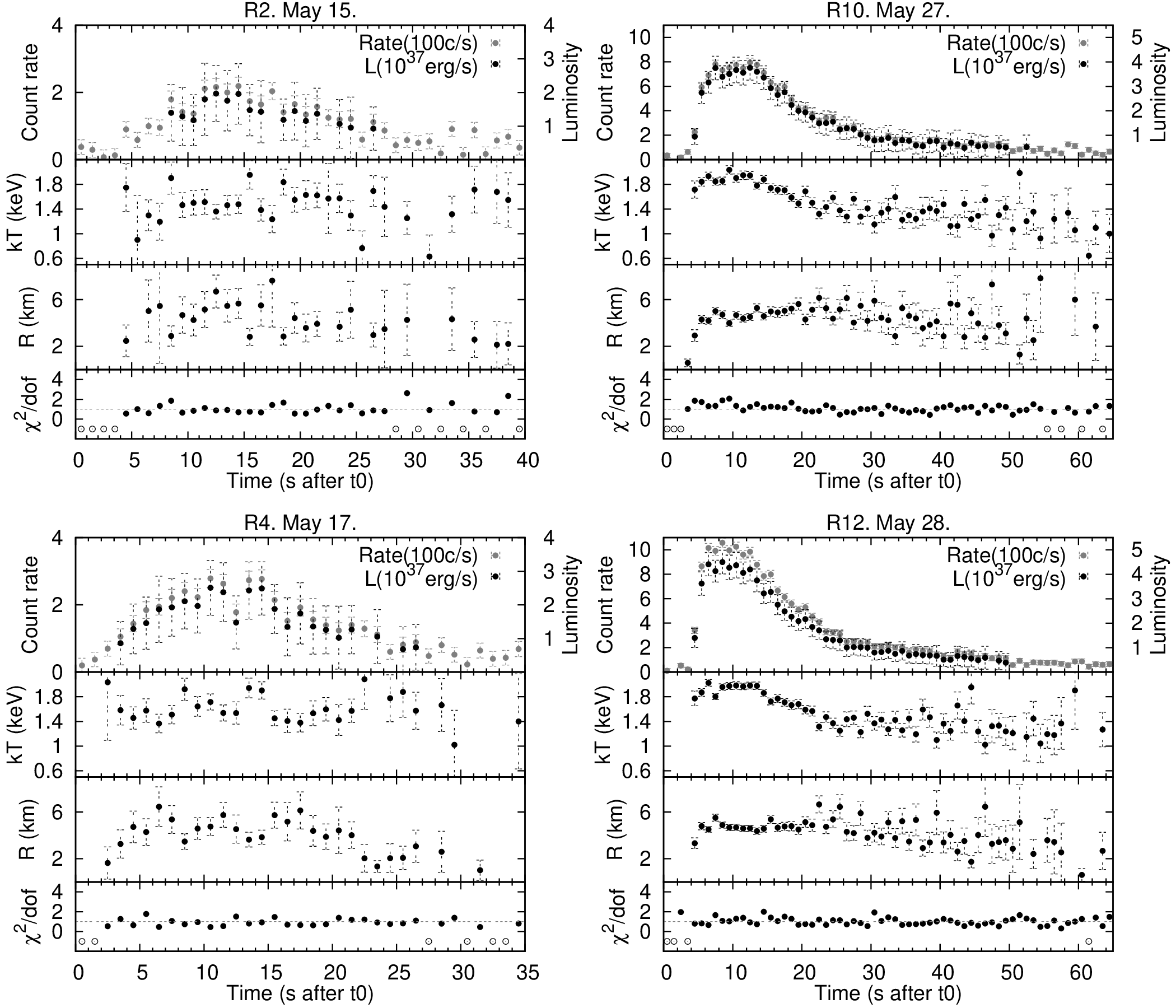}}}
  \caption{Time-resolved spectroscopy of two representative ``early bursts'' (R2 and R4; {\it left}) and prototypical type I X-ray bursts (R10 and R12; {\it right} see Sec.~\ref{sec:rxte}). Panels show, from top to bottom: net 2.5-25.0~keV count rate (all active PCUs; see Table~\ref{table:rbursts}); black-body temperature, black-body radius (at 7.8~kpc) and reduced $\chi^2$ (open circles mark spectra with less than 50 counts, not fitted).}
    \label{fig:btspec}
%\epsscale{1.0}
% \end{center}
\end{figure}

We present in Table~\ref{table:rbursts} the properties of all X-ray
bursts detected by {\it RXTE}. Detailed inspection of these properties
reveals two clearly distinct flavors: bursts R1-R5 (the ``early
bursts'') have long rise times (7.3--10.2~s), moderate energy output
(total energy of [1.3--3.7]$\times$10$^{38}$erg), and show
approximately symmetric profiles. On the other hand, bursts R6-R12
feature shorter rise times (1.8--5.3~s), are more energetic
([5.3--11]$\times$10$^{38}$erg) and present prototypical type I X-ray
burst (fast-rise-exponential-decay, FRED-like) light~curves. We show
in Figure~\ref{fig:btspec} the spectral evolution of bursts R2, R4,
R10 and R12, two representative examples of each class. The peak
luminosities of the early bursts were systematically lower than those
of bursts R6-R12, and the persistent pre-burst luminosities were
higher on average in the early bursts than in bursts R6-R12, although
with overlaps (see Table~\ref{table:rbursts}). Besides the
above-mentioned FRED-like profile, bursts R6-R12 all show clear
cooling trends along their decays (see Figure~\ref{fig:btspec}), and
we therefore classify them unequivocally as type I (thermonuclear)
X-ray bursts. The early bursts (R1-R5) showed little or no signs of
cooling along the tail, with black body temperature approximately
constant in the range $\sim$1.2--1.8~keV; we discuss their possible
origins in Section~\ref{sec:discussion}, and argue that they are most
likely also of thermonuclear nature.

Each burst was searched for burst oscillations in the range
10--2048~Hz. We searched for signals using the entire burst and
shorter (4s) time windows. We found no events where the significance
exceeded 3$\sigma$ after accounting for numbers of trials.

\subsection{Swift}
\label{sec:swift}

We asked for {\it Swift} ToO observations of the source to confirm the
location of the bursts. Four observations were performed between 2010
May 27--29, and no bursts were detected during the first
observation \citep{Papitto10a}. We found an X-ray burst in the second
observation (\citealt{Linares10b}; see also \citealt{Papitto10b}). We
report the detection of a third X-ray burst with {\it Swift}-XRT in a
photon-counting (PC) mode observation that took place on May 29th (S3,
see below).

We processed the XRT data using {\it xrtpipeline} with standard
quality cuts. Exposure maps and ancillary response files were created
using {\it xrtexpomap} and {\it xrtmkarf}, respectively. The
latest response matrix files (v11) were taken from the calibration
database. Spectra were grouped to contain a minimum of 10 and 20
photons per bin for burst and persistent spectra, respectively. We
extracted source events from a 40x40 pixel box for the WT data and
from a 9-40 pixel annulus for the PC data.

Table~\ref{table:rbursts} presents the main properties of the three
{\it Swift} bursts (S1-S3). The analysis of S3 is affected by severe
pile-up of the XRT in PC mode at the collected count rates (more than
$\sim$15 c/s at the peak). The burst profile is also typical of type I
X-ray bursts, and we consider it likely that this is a type I X-ray
burst. Due to the low count rates collected by the XRT the spectral
information that can be extracted from S1-S2 is limited. However, by
extracting three spectra along the tail of bursts S1 and S2 (and using
a pre-burst spectrum as background; Sec.~\ref{sec:rxte}) we are able
to constrain the black-body temperature and we find evidence of
cooling in both bursts. The light~curves show typical FRED profiles
and the temperatures and black body radii that we find are fully
consistent with those measured by {\it RXTE}. We therefore identify
S1-S2 as type~I (thermonuclear) X-ray bursts.

\begin{figure}[h!]
  \begin{center}
  \resizebox{0.8\columnwidth}{!}{\rotatebox{-90}{\includegraphics[]{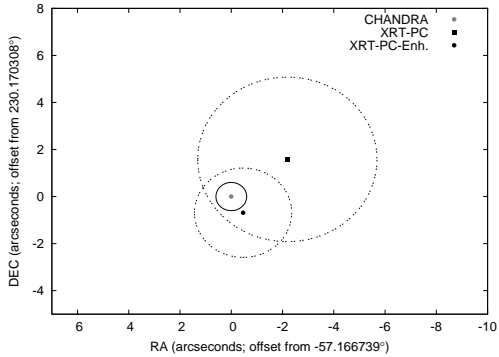}}}
  \caption{{\it Swift}-XRT location of Cir~X-1 during its May 2010 bursting period. The grey filled circle and solid line show the {\it Chandra} position \citep[0.6~arcsec error radius]{Iaria08}. The filled black square and circle show the {\it Swift}-XRT PC mode position and UVOT-enhanced position \citep{Goad07}, and the dashed circles show their 3.5 and 1.9~arsec radius respective error circles. We detect one X-ray burst during the PC mode observation (S3).}
    \label{fig:locate}
%\epsscale{1.0}
 \end{center}
\end{figure}

We fitted the persistent spectra of the four {\it Swift} observations
in order to measure the flux evolution around the bursts. We show the
resulting 2--10~keV absorbed flux in Figure~\ref{fig:lc2} (right). No
other sources were detected in the XRT FoV during the May 29 PC mode
observation \citep{Papitto10b}. We obtain, using {\it xrtcentroid},
the following position: R.A.=15$^h$20$^m$40.73$^s$,
DEC=-57$^\circ$09$'$58.7$''$ (J2000.0), with a 90\% confidence error
radius of 3.5~arcsec. We also obtain a UVOT-enhanced XRT position
\citep{Evans09,Goad07} of: R.A.=15$^h$20$^m$40.84$^s$,
DEC=-57$^\circ$10$'$00.9$''$ (J2000.0; 1.9~arcsec 90\% confidence
error radius). As shown in Figure~\ref{fig:locate}, both positions are
consistent with the {\it Chandra} position given by
\citet{Iaria08}. 
The $\sim$345  photons collected during burst S3 cluster around Cir~X-1,
with 90\% of the photons within 6 arcmin of the {\it Chandra} position.
We therefore conclude
that Cir~X-1 is the origin of bursts S1-S3 and, in all likelihood,
R1-R12.

%Given the detection of a burst in the same May 29
%observation (S3), the reported {\it Swift} position shows that the
%burster is located within 1.9~arcsec of Cir~X-1. 

%path is: /scratch/mlinares/rxte/CirX-1/RT/BURSTS/

\section{Discussion}
\label{sec:discussion}

We have reported the first X-ray bursts detected from Cir~X-1 since
1985. A total of 15 bursts were recorded during May 15-30 2010, when
source flux was in the range 0.3--0.01~Crab. The early bursts were
short, low-luminosity events with recurrence times as short as 20
minutes and no evidence for cooling.  Later bursts were longer and
brighter, with cooling blackbody tails that clearly identify them as
thermonuclear bursts. The general trend in persistent flux was
downwards, but with fluctuations (Fig.~\ref{fig:lc2} left,
Table~\ref{table:rbursts}).

There are strong similarities between the 2010 bursts and those seen
in 1984-5.  \citet{Tennant86a}, in December 1984, observed 8 low
luminosity bursts with wait times in the range 27--47 minutes and
marginal evidence for cooling in the burst tails. In August 1985, at a
similar persistent flux, the same authors observed 3 bursts with
typical type I burst properties \citep{Tennant86b}. Peak fluxes and
temperatures were similar to those we see in 2010. The fact that the
earlier bursts are so similar to those seen in 2010 gives us
confidence that the 1984-5 bursts genuinely originated from Cir~X-1.
Persistent flux when the 1984-5 bursts were observed was
$\sim$0.1~Crab \citep{Parkinson03}.

Our results confirm that Cir~X-1 is a neutron star rather than a black
hole (the arcsec location of an X-ray burst from Cir~X-1 is
conclusive), but also pose two major questions. Firstly, why have no
bursts been seen from this source in the intervening 25 years? And
secondly, why are the burst properties so diverse? In the rest of this
Section we will address these two issues.

%It is clear from the long-term lightcurve of this source that 

From 1985 to 2003 the source was observed almost exclusively at
persistent luminosities much higher than those measured during May
2010 (\citealt{Parkinson03}; Figure~\ref{fig:lc2}-left in this
paper). High accretion rates suppress thermonuclear instabilities
\citep[e.g.][]{Cornelisse03}, which may explain the fact that no
bursts were observed in this period despite extensive time on
source. From April 2003 (MJD $\sim$52750), however, the average source
flux decayed below $\sim$0.3~Crab.  Between April 2003 and April 2010,
Cir~X-1 was observed for more than 600~ks by the {\it RXTE}-PCA in the
0.01--0.3~Crab flux range, the same flux measured during May 2010. No
X-ray bursts were detected during this period, even though accretion
rate was similar to that seen during the May 2010 re-brightening,
judging from the X-ray flux. For recurrence times similar to those
observed in May 2010, the chances of detecting bursts in 600~ks would
be very high, if they were present. To illustrate this, one can
compare the average burst rate during our May 2010 observations,
0.6~hr$^{-1}$, to the 95\% upper limit \citep{Gehrels86} on the
average burst rate between April 2003 and April 2010,
$<$0.018~hr$^{-1}$ (given the non-detection in 600~ks). Such extreme
difference in bursting rate at a similar flux level suggests that an
additional parameter, other than instantaneous accretion rate, must be
invoked to explain the ``bursting'' and ``non-bursting'' regimes (see
below).

With regard to the variability in burst properties, we consider three
possible scenarios. The first, which was also discussed by
\citet{Tennant86a}, is that the early bursts are type II bursts,
powered by accretion instabilities\citep[see][and references therein
]{Lewin96}. We consider this possibility unlikely, for the following
reasons. (1) Rise times for the Cir~X-1 early bursts are slower and
(where they can be calculated) $\alpha$ values much higher than those
for typical type II bursts; (2) There is no evidence for low frequency
QPOs, often observed with type II bursts. For the triple burst
observation, on May 20, we estimate 3-$\sigma$ upper limits on the
fractional rms amplitude of a QPO between $\sim$2 and $\sim$8~Hz
(1.8~Hz FWHM) of 3\% (between bursts) and 7\% (during bursts), lower
than the typical fractional amplitudes reported by \citet{Lubin92} and
\citet{Dotani90}. (3) Accretion instabilities would be expected to
recur at the same accretion rate. The fact that bursts were not
observed from 2003-2010 even when the source was at a similar
persistent flux would argue against this. And (4), the fact that the
early bursts were observed only three days before the confirmed type I
bursts, after 25 years with no bursts detected, suggests a common
origin.

A second possibility is that we are observing the transition from
stable thermonuclear burning of Helium at high accretion rates to
unstable burning at lower rates \citep{Bildsten98}. This transition
region permits some interesting behavior, as outlined by
\citet{Heger07}. As accretion rate falls, these authors showed that
one should expect first marginally stable quasi-periodic burning (mHz
QPOs), then low-luminosity short recurrence time bursts, before the
eventual establishment of brighter bursts with longer recurrence
times. If the brightest fluxes observed from Cir~X-1 correspond to
Eddington rate accretion then this scenario is plausible, although
there is no evidence for mHz QPOs in the 2010 data. This scenario
might also explain the 1984-5 observations, where, interestingly,
there is some evidence in the light~curve for variability on
$\sim$1000~s timescales (see Figure 1 of \citealt{Tennant86a}). If
this is the case then Cir~X-1 would be a valuable probe of this
transition, since very short recurrence time bursts at high accretion
rates are extremely rare \citep{Keek10}.

The third possibility is that the accretion rate of Cir~X-1 is lower,
and that we are seeing short recurrence time thermonuclear bursts
similar to those seen in other sources, interspersed with more regular
bursts \citep{Keek10}. In the general population of bursting sources,
bursts with recurrence times of less than 40 minutes are found only at
accretion rates less than 5\% of the Eddington rate. If this is the
case in Cir~X-1 then the peak brightness observed in the RXTE era
would be substantially less than Eddington. If Cir~X-1 has never been
accreting close to the Eddington limit, this would have implications
for the determination of the distance, favouring a lower distance as
proposed by \citet{Iaria05}. It would however be difficult to explain
Z-source behavior from a source with such low accretion rate.

If the short bursts are thermonuclear in origin, then the question of
why they were not seen by RXTE on earlier occasions (from 2003
onwards) remains. One very intriguing possibility, that would lend
support to the high accretion rate scenario outlined above, is that
the heating associated with a prolonged period of accretion has been
acting to stabilize the burning process. Sustained accretion can heat
the neutron star crust, which then cools over an extended period,
maintaining a high temperature in the ocean even after accretion rate
has fallen \citep{haensel1990,Brown98,Brown09}. Additional flux from
below the burning layer stabilizes He burning
\citep{Bildsten95,Bildsten98,Cumming04}. The re-brightening of Cir~X-1
in 2010 took place after a prolonged period ($\sim$2 years) of very
low accretion rate, much longer than had been experienced earlier in
{\it RXTE}'s lifetime.  This may be the first time that the crust and
ocean have been able to cool sufficiently for burning to be unstable
at the observed accretion rates. The accretion history in the few
years prior to the 1984-5 bursting episode is unfortunately not
available to cross-check this \citep{Parkinson03}, but it is an
interesting possibility.

\textbf{Acknowledgments:} 

This work made use of data supplied by the ASM/RXTE team, by RIKEN,
JAXA and the MAXI team and by the Swift SDC at the University of
Leicester. ML acknowledges support from the Netherlands Organization
for Scientific Research Rubicon fellowship.

.

%\clearpage

%\newpage 
%\clearpage

%\bibliographystyle{aa}
%\bibliography{/home/mlinares/papers/biblio.bib}
%\bibliography{biblio.bib}

\begin{thebibliography}{48}
\expandafter\ifx\csname natexlab\endcsname\relax\def\natexlab#1{#1}\fi
\expandafter\ifx\csname url\endcsname\relax
  \def\url#1{{\tt #1}}\fi
\expandafter\ifx\csname urlprefix\endcsname\relax\def\urlprefix{URL }\fi

\bibitem[{{Anders} \& {Ebihara}(1982)}]{Anders82}
{Anders} E., {Ebihara} M., 1982, \gca, 46, 2363

\bibitem[{{Bildsten}(1995)}]{Bildsten95}
{Bildsten} L., 1995, \apj, 438, 852

\bibitem[{{Bildsten}(1998)}]{Bildsten98}
{Bildsten} L., 1998, In: {R.~Buccheri et~al.} (ed.) NATO
  ASIC Proc. 515: The Many Faces of Neutron Stars., 419--+

\bibitem[{{Boutloukos} et~al.(2006){Boutloukos}, {van der Klis}, {Altamirano}
  et~al.}]{Boutloukos06}
{Boutloukos} S., et~al., 2006, \apj, 653,
  1435

\bibitem[{{Brown} \& {Cumming}(2009)}]{Brown09}
{Brown} E.F., {Cumming} A., 2009, \apj, 698, 1020

\bibitem[{{Brown} et~al.(1998){Brown}, {Bildsten}, \& {Rutledge}}]{Brown98}
{Brown} E.F., et~al., 1998, \apjl, 504, L95+

\bibitem[{{Cornelisse} et~al.(2003){Cornelisse}, {in't Zand}, {Verbunt}
  et~al.}]{Cornelisse03}
{Cornelisse} R., et~al., 2003, \aap, 405,
  1033

\bibitem[{Cumming \& Macbeth(2004)}]{Cumming04}
Cumming A., Macbeth J., Mar 2004, \apj, 603, L37

\bibitem[{{Dotani} et~al.(1990){Dotani}, {Mitsuda}, {Inoue} et~al.}]{Dotani90}
{Dotani} T., et~al., 1990, \apj, 350, 395

\bibitem[{{Evans} et~al.(2009){Evans}, {Beardmore}, {Page} et~al.}]{Evans09}
{Evans} P.A., et~al., 2009, \mnras, 397, 1177

\bibitem[{{Galloway} et~al.(2008){Galloway}, {Muno}, {Hartman}, {Psaltis}, \&
  {Chakrabarty}}]{Galloway08}
{Galloway} D.K., et~al., 2008, \apjs, 179, 360

\bibitem[{{Gehrels}(1986)}]{Gehrels86}
{Gehrels} N., 1986, \apj, 303, 336

\bibitem[{{Goad} et~al.(2007){Goad}, {Tyler}, {Beardmore} et~al.}]{Goad07}
{Goad} M.R., et~al., 2007, \aap, 476, 1401

\bibitem[{{Haensel} \& {Zdunik}(1990)}]{haensel1990}
{Haensel} P., {Zdunik} J.L., Jan. 1990, \aap, 227, 431

\bibitem[{{Heger} et~al.(2007){Heger}, {Cumming}, \& {Woosley}}]{Heger07}
{Heger} A., et~al., 2007, \apj, 665, 1311

\bibitem[{{Iaria} et~al.(2001){Iaria}, {Di Salvo}, {Burderi}, \&
  {Robba}}]{Iaria01}
{Iaria} R., et~al., 2001, \apj, 561, 321

\bibitem[{{Iaria} et~al.(2005){Iaria}, {Span{\`o}}, {Di Salvo}
  et~al.}]{Iaria05}
{Iaria} R., et~al., 2005, \apj, 619, 503

\bibitem[{{Iaria} et~al.(2008){Iaria}, {D'A{\'{\i}}}, {Lavagetto}
  et~al.}]{Iaria08}
{Iaria} R., et~al., 2008, \apj, 673, 1033

\bibitem[{{Jahoda} et~al.(1996){Jahoda}, {Swank}, {Giles} et~al.}]{Jahoda96}
{Jahoda} K., et~al., 1996, In: {Siegmund} O.H.,
  {Gummin} M.A. (eds.) SPIE Conference Series, vol. 2808, 59--70

\bibitem[{{Jonker} \& {Nelemans}(2004)}]{Jonker04}
{Jonker} P.G., {Nelemans} G., 2004, \mnras, 354, 355

\bibitem[{{Jonker} et~al.(2007){Jonker}, {Nelemans}, \& {Bassa}}]{Jonker07}
{Jonker} P.G., et~al., 2007, \mnras, 374, 999

\bibitem[{{Kaluzienski} et~al.(1976){Kaluzienski}, {Holt}, {Boldt}, \&
  {Serlemitsos}}]{Kaluzienski76}
{Kaluzienski} L.J., et~al., 1976, \apjl, 208, L71

\bibitem[{{Keek} et~al.(2010){Keek}, {Galloway}, {in 't Zand}, \&
  {Heger}}]{Keek10}
{Keek} L., et~al., 2010, ArXiv e-prints

\bibitem[{{Kuulkers} et~al.(2002){Kuulkers}, {Homan}, {van der Klis}, {Lewin},
  \& {M{\'e}ndez}}]{Kuulkers02}
{Kuulkers} E., et~al., 2002, \aap, 382, 947

\bibitem[{{Levine} et~al.(1996){Levine}, {Bradt}, {Cui} et~al.}]{Levine96}
{Levine} A.M., et~al., 1996, \apjl, 469, L33+

\bibitem[{{Lewin} et~al.(1993){Lewin}, {van Paradijs}, \& {Taam}}]{Lewin93}
{Lewin} W.H.G., et~al., 1993, Space Science Reviews, 62, 223

\bibitem[{{Lewin} et~al.(1996){Lewin}, {Rutledge}, {Kommers}, {van Paradijs},
  \& {Kouveliotou}}]{Lewin96}
{Lewin} W.H.G., et~al., 1996, \apjl, 462, L39+

\bibitem[{{Linares} et~al.(2010{\natexlab{a}}){Linares}, {Soleri}, {Altamirano}
  et~al.}]{Linares10b}
{Linares} M., et~al., 2010{\natexlab{a}}, The Astronomer's Telegram, 2651, 1

\bibitem[{{Linares} et~al.(2010{\natexlab{b}}){Linares}, {Soleri}, {Watts}
  et~al.}]{Linares10a}
{Linares} M., et~al., 2010{\natexlab{b}}, The Astronomer's Telegram, 2643, 1

\bibitem[{{Lubin} et~al.(1992){Lubin}, {Lewin}, {Rutledge} et~al.}]{Lubin92}
{Lubin} L.M., et~al., 1992, \mnras, 258, 759

\bibitem[{{Margon} et~al.(1971){Margon}, {Lampton}, {Bowyer}, \&
  {Cruddace}}]{Margon71}
{Margon} B., et~al., 1971, \apjl, 169, L23+

\bibitem[{{Matsuoka} et~al.(2009){Matsuoka}, {Kawasaki}, {Ueno}
  et~al.}]{Matsuoka09}
{Matsuoka} M., et~al., 2009, \pasj, 61, 999

\bibitem[{{Murdin} et~al.(1980){Murdin}, {Jauncey}, {Lerche} et~al.}]{Murdin80}
{Murdin} P., et~al., 1980, \aap, 87, 292

\bibitem[{{Nakajima} et~al.(2010){Nakajima}, {Matsuoka}, {Kawasaki}
  et~al.}]{Nakajima10}
{Nakajima} M., et~al., 2010, The Astronomer's
  Telegram, 2608, 1

\bibitem[{{Nicolson}(2007)}]{Nicolson07}
{Nicolson} G.D., 2007, The Astronomer's Telegram, 985, 1

\bibitem[{{Oosterbroek} et~al.(1995){Oosterbroek}, {van der Klis}, {Kuulkers},
  {van Paradijs}, \& {Lewin}}]{Oosterbroek95}
{Oosterbroek} T., et~al., 1995, \aap, 297, 141

\bibitem[{{Papitto} et~al.(2010{\natexlab{a}}){Papitto}, {Bozzo}, {D'Ai}
  et~al.}]{Papitto10b}
{Papitto} A., et~al., 2010{\natexlab{a}}, The Astronomer's Telegram, 2653, 1

\bibitem[{{Papitto} et~al.(2010{\natexlab{b}}){Papitto}, {D'Ai}, {Bozzo},
  {Iaria}, \& {Di Salvo}}]{Papitto10a}
{Papitto} A., et~al., 2010{\natexlab{b}}, The Astronomer's Telegram, 2650, 1

\bibitem[{{Parkinson} et~al.(2003){Parkinson}, {Tournear}, {Bloom}
  et~al.}]{Parkinson03}
{Parkinson} P.M.S., et~al., 2003, \apj, 595, 333

\bibitem[{{Schulz} et~al.(2008){Schulz}, {Kallman}, {Galloway}, \&
  {Brandt}}]{Schulz08}
{Schulz} N.S., et~al., 2008, \apj, 672, 1091

\bibitem[{{Shirey} et~al.(1998){Shirey}, {Bradt}, {Levine}, \&
  {Morgan}}]{Shirey98}
{Shirey} R.E., et~al., 1998, \apj, 506, 374

\bibitem[{{Soleri} et~al.(2009){Soleri}, {Tudose}, {Fender}, {van der Klis}, \&
  {Jonker}}]{Soleri09}
{Soleri} P., et~al., 2009, \mnras, 399, 453

\bibitem[{{Strohmayer} \& {Bildsten}(2006)}]{Strohmayer06}
{Strohmayer} T., {Bildsten} L., 2006, {New views of thermonuclear bursts},
  113--156

\bibitem[{{Tennant} et~al.(1986{\natexlab{a}}){Tennant}, {Fabian}, \&
  {Shafer}}]{Tennant86a}
{Tennant} A.F., et~al., 1986{\natexlab{a}}, \mnras, 219, 871

\bibitem[{{Tennant} et~al.(1986{\natexlab{b}}){Tennant}, {Fabian}, \&
  {Shafer}}]{Tennant86b}
{Tennant} A.F., et~al., 1986{\natexlab{b}}, \mnras, 221, 27P

\bibitem[{{Toor}(1977)}]{Toor77}
{Toor} A., 1977, \apjl, 215, L57

\bibitem[{{van der Klis}(1994)}]{Klis94}
{van der Klis} M., 1994, \apjs, 92, 511

\bibitem[{{van der Klis}(2006)}]{vanderKlis06}
{van der Klis} M., 2006, in "Compact Stellar X-ray Sources", ed. W. H. G. Lewin
  \& M. van der Klis (Cambridge Univ. Press) (astro-ph/0410551), 39--112

\end{thebibliography}

\end{document}